\documentclass[aps,pre,showpacs,reprint]{revtex4-1}

\usepackage{amssymb,amsmath}

\usepackage{bm}
\usepackage{color}
\usepackage{graphicx}

\newcommand\beq{\begin{equation}}
\newcommand\eeq{\end{equation}}
\newcommand\beqa{\begin{eqnarray}}
\newcommand\eeqa{\end{eqnarray}}
\newcommand{\nn}{\nonumber\\}

\newcommand{\at}{\widetilde{\alpha}}
\newcommand{\bt}{\widetilde{\beta}}
\newcommand{\al}{\alpha}
\newcommand{\be}{\beta}

\newcommand{\q}{\kappa}

\newcommand{\JJ}{\bm{\Delta}_{12}}
\newcommand{\JJb}{\bar{\bm{\Delta}}_{12}}

\begin{document}

\title{Role of roughness on the hydrodynamic homogeneous base state of inelastic spheres}
\author{Francisco Vega Reyes}
\email{fvega@unex.es}

\author{Andr\'es Santos}
\email{andres@unex.es}
\affiliation{Departamento de F\'isica, Universidad de Extremadura, 06071 Badajoz, Spain}

\author{Gilberto M.  Kremer}
\email{kremer@fisica.ufpr.br}
\affiliation{Departamento de F\'{\i}sica, Universidade Federal do Paran\'a, Curitiba, Brazil}


\begin{abstract}
A gas of inelastic rough spheres admits a spatially homogeneous base state which turns into a hydrodynamic state after a finite relaxation time. We show that this relaxation time is hardly dependent on the degree of inelasticity but increases dramatically with decreasing roughness.  An accurate description of translational-rotational velocity correlations at all times is also provided. At a given inelasticity, the roughness parameter can be tuned to produce a huge distortion from the Maxwellian distribution function. The results are obtained from a Grad-like solution of the Boltzmann--Enskog equation complemented by Monte Carlo {and molecular dynamics} simulations.

\end{abstract}

\date{\today}
\maketitle

The extension of statistical and fluid mechanics concepts to fluidized granular matter systems has allowed for a better understanding of the discrete-to-continuum  description of matter  by placing it into a more general theoretical framework \cite{JNB96a,D01,G03,BP04}.

Granular systems may exist in fluidized states at densities low enough to make a description by means of the (inelastic) Boltzmann and Enskog equations \cite{D01,G03,GNB05,GNB05b,BB12}  possible. In that context, an immediate question {arises}: {Do the inelastic versions of these kinetic equations} support an accurate hydrodynamic description for granular gases as the elastic one \cite{CC70} does for molecular gases? A variety of interesting kinetic theory studies have successfully modeled granular gas dynamics, proving additionally that granular gases may admit a hydrodynamic description \cite{BDKS98,D01,G03,BP04,S08,VSG10}. However, most granular transport theories do not take into account the effects of {particle roughness, which} is inherently present in all real granular systems \cite{GS95,KH04}, {or do it in the quasismooth regime \cite{GNB05,GNB05b}. Thus,} the debate on the limits of applicability of granular hydrodynamics is still not closed.

The existence of a hydrodynamic {regime} relies on scale separation \cite{C00a}, i.e., individual particle (microscopic) dynamics variations (both in time and space) should be much shorter than those for the (macroscopic) average fields \cite{CC70,C00a,D01}. For molecular fluids, {hydrodynamic states} exist if the system is not subject to large gradients from the boundaries. However, for granular gases, even if no gradients are applied at all, the inelastic cooling sets an inherent decay time rate for the kinetic energy which is not necessarily slow compared to the characteristic microscopic time. This {makes} the proof of existence of a hydrodynamic solution in granular gases be not trivial \cite{BD05}, even for homogeneous states.  {The} more realistic case of rough spheres seems to be much more complex \cite{GS95}. For instance, energy nonequipartition \cite{LHMZ98,HHZ00}, non-Maxwellian behavior \cite{GNB05,GNB05b,BB12}, and correlations between translational and angular velocities  \cite{BPKZ07,GA08,KBPZ09}  {appear}.

Taking the homogeneous cooling state (HCS) \cite{BP04} as the base state for hydrodynamics, we {address  in this Rapid Communication questions such as the following: Is the ability of a homogeneous gas of rough spheres to reach a hydrodynamic state  related to the degrees of inelasticity and/or roughness? How does the degree of roughness affect the aging time needed to reach the hydrodynamic HCS state? Is the HCS marginal probability distribution of angular velocities close to a Maxwellian? For this,  {both theoretical and simulational routes are followed}. We develop a perturbative, Grad-like solution of the Boltzmann--Enskog (BE) equation that takes into account the effects of translational-rotational velocity correlations and non-Maxwellian {features of the velocity distribution function}.  Moreover, we confirm our theory results by {numerical solutions of the BE equation by means of} the direct simulation Monte Carlo (DSMC) method \cite{BP04}. {In order to check that eventual violations of molecular chaos are not relevant, we carry out additional molecular dynamics (MD) simulations, using and event-driven algorithm \cite{VGS11,note_13_11_1}.}

{The BE equation for a homogeneous state reads}
\begin{equation}
  \label{BE}
  \frac{\partial f}{\partial t}={\sigma^2\chi K[\mathbf{v}, \bm{\omega}|f]},
\end{equation}
where $f(\mathbf{v}, \bm{\omega},t)$ is the one-body distribution function, $\mathbf{v}$ and $\bm{\omega}$ being the particle translational and angular velocities, respectively, {$\sigma$ is the sphere diameter, $\chi$ is the pair correlation function at contact (Enskog factor), which accounts for finite-density effects \cite{HM06}, and ${\sigma^2\chi K\equiv J}$ is the usual collision operator for inelastic and rough hard spheres \cite{K10a,note_13_09_1}. The collision rule involves the normal ($\alpha$) and tangential ($\beta$) coefficients of restitution \cite{SKS11,K10a}. While $\alpha$ ranges from $0$ (perfectly inelastic) to $1$ (perfectly elastic), $\beta$ ranges from $-1$ (perfectly smooth) to $1$ (perfectly rough)}. A more detailed description on the mechanics of collisions of rough hard spheres may be found elsewhere \cite{FLCA94,GS95,BPKZ07,SKS11}.

{While the two-parameter $(\alpha,\beta)$ model neglects sliding effects that can be relevant in
grazing collisions \cite{FLCA94}, a more sophisticated collision model with a  Coulomb friction constant \cite{W93} may hinder the
possibility of analytical treatments outside of the quasielastic and/or quasismooth limits \cite{GNB05,JZ02}.
Moreover, the $(\alpha,\beta)$ model still captures the basic features of the collision process and
of the hydrodynamic issue \cite{BPKZ07,GA08,MDHEH13} without compromising its physical content as a granular fluid model.}

The translational and rotational temperatures are defined in the usual way \cite{BPKZ07,SKS11} as
$T_t=\frac{m}{3}\langle(\mathbf{v}-\mathbf{u}_f)^2\rangle$ and $T_r=\frac{I}{3}\langle\omega^2\rangle$, respectively,
where $m$ and $I$ are the mass and moment of inertia, respectively,  of the particles, and $\mathbf{u}_f=\langle \mathbf{v}\rangle$ is the flow velocity. The total temperature $T=\frac{1}{2}(T_t+T_r)$ decays monotonically with time (unless $\alpha=\beta^2=1$). Since both $\mathbf{u}_f$ and the density $n$ are constant in homogeneous states, $T$ is the only relevant hydrodynamic quantity of the system. Thus, if a hydrodynamic regime {does} exist, the whole temporal dependence of $f$ {must} occur through a dependence on $T$ \cite{vNE98}.

The investigation of this scenario {calls for} the introduction of the reduced translational and angular velocities $\mathbf{c}(t)\equiv(\mathbf{v}-\mathbf{u}_f)/\sqrt{2T_t(t)/m}$, $\mathbf{w}(t)\equiv\bm{\omega}/\sqrt{2T_r(t)/I}$, and the reduced distribution function $\phi(\mathbf{c},\mathbf{w},t)\equiv {n}^{-1}\left[4T_t(t)T_r(t)/m I\right]^{3/2}f(\mathbf{v},\bm{\omega},t)$. In terms of these reduced quantities, the evolution equation for the temperature ratio $\theta(t)\equiv T_r(t)/T_t(t)$ and the BE equation \eqref{BE} become
\beq
  \label{partialtheta}
  \partial_\tau\ln\theta=-\frac{2}{3}\left(\mu_{02}^{(0)}-\mu_{20}^{(0)}\right),
\eeq
\beq
  \label{rBE}
\partial_\tau\phi+\frac{\mu_{20}^{(0)}}{3}\frac{\partial}{\partial\mathbf{c}}\cdot\left(\mathbf{c}\phi\right)
+\frac{\mu_{02}^{(0)}}{3}\frac{\partial}{\partial\mathbf{w}}\cdot\left(\mathbf{w}\phi\right)= {\mathcal{J}[\mathbf{c},\mathbf{w}|\phi]},
\eeq
where $\partial_\tau\equiv [\nu(t)]^{-1}\partial_t$ is a time derivative scaled by the effective collision frequency $\nu(t)=2n\sigma^2\chi \sqrt{\pi T_t(t)/m}$, $\mathcal{J}\equiv \nu^{-1} {n}^{-1}\left(4T_tT_r/m I\right)^{3/2} J$ is the reduced collision operator \cite{note_13_09_1}, and
\beq
  \label{mupq}
\mu_{pq}^{(r)}\equiv-\int d\mathbf{c}\int d\mathbf{w}\,
c^pw^q
(\mathbf{c}\cdot\mathbf{w})^r
{\mathcal{J}[\mathbf{c},\mathbf{w}|\phi]}
\eeq
are reduced collisional moments.
Taking moments on both sides of Eq.\ \eqref{rBE} we get the equations
\begin{equation}
\partial_\tau \ln M_{pq}^{(r)}-\frac{p+r}{3}\mu_{20}^{(0)}-\frac{q+r}{3}\mu_{02}^{(0)}=-{\mu_{pq}^{(r)}}/{M_{pq}^{(r)}},
\label{partialpq}
\end{equation}
where $M_{pq}^{(r)}\equiv \langle c^pw^q
(\mathbf{c}\cdot\mathbf{w})^r\rangle$.
{If a hydrodynamic description applies,  it is expected that, after a certain transient period (kinetic stage), the system reaches an asymptotic regime (hydrodynamic stage)} where $\theta$ and $\phi(\mathbf{c},\mathbf{w})$ (or, equivalently, its moments  $M_{pq}^{(r)}$) become independent of time.

In isotropic conditions,  $\phi(\mathbf{c},\mathbf{w})$ is
actually a function of the three scalar quantities $c^2=\mathbf{c}\cdot\mathbf{c}$, $w^2=\mathbf{w}\cdot\mathbf{w}$, and $(\mathbf{c}\cdot \mathbf{w})^2$. As a consequence, one can formally represent the ratio $\phi(\mathbf{c},\mathbf{w})/\phi_M(\mathbf{c},\mathbf{w})$, where $\phi_M(\mathbf{c},\mathbf{w})=\pi^{-3}e^{-c^2-w^2}$ is the (reduced) {two-temperature} Maxwellian distribution, as an infinite series of \emph{polynomials} in $c^2$, $w^2$, and $(\mathbf{c}\cdot \mathbf{w})^2$:
\beq
\phi(\mathbf{c},\mathbf{w})=\phi_M(\mathbf{c},\mathbf{w})\sum_{j=0}^\infty\sum_{k=0}^\infty\sum_{\ell=0}^\infty a_{jk}^{(\ell)}\Psi_{jk}^{(\ell)}(\mathbf{c},\mathbf{w}),
\label{phi}
\eeq
where
$\Psi_{jk}^{(\ell)}(\mathbf{c},\mathbf{w})=L_j^{(2\ell+\frac{1}{2})}(c^2) L_k^{(2\ell+\frac{1}{2})}(w^2)\left(c^2w^2\right)^\ell
P_{2\ell}(u)$ is a polynomial of total degree   $2(j+k+2\ell)$ in velocity.
Here, $L_j^{(2\ell+\frac{1}{2})}(x)$ and $P_{2\ell}(x)$ are Laguerre  and Legendre polynomials, respectively, and
$u\equiv (\mathbf{c}\cdot \mathbf{w})/cw$ is the cosine of the angle made by $\mathbf{v}$ and $\bm{\omega}$. The set of polynomials $\{\Psi_{jk}^{(\ell)}\}$ is a complete orthogonal basis for the solution of Eq.\ \eqref{rBE} \cite{note_13_09_1}. The expansion coefficients $a_{jk}^{(\ell)}\propto \langle \Psi_{jk}^{(\ell)}\rangle$ are linear combinations of the moments $M_{pq}^{(r)}$ with $p,q,r=\text{even}$ and $p+q+2r\leq 2(j+k+2\ell)$. By normalization, $a_{00}^{(0)}=1$, $a_{10}^{(0)}=a_{01}^{(0)}=0$, so the first nontrivial coefficients are those of degree four, namely, the {fourth-degree} \emph{cumulants}
\begin{subequations}
\begin{equation}
a_{20}^{(0)}=\frac{4}{15}\langle c^4\rangle-1, \quad a_{02}^{(0)}=\frac{4}{15}\langle w^4\rangle-1,
\label{a20}
\end{equation}
\begin{equation}
 a_{11}^{(0)}=\frac{4}{9}\langle c^2w^2\rangle-1,\quad
a_{00}^{(1)}=\frac{8}{15}\left[\langle (\mathbf{c}\cdot\mathbf{w})^2\rangle-\frac{1}{3}\langle c^2w^2\rangle\right].
\label{b}
\end{equation}
\label{ajkl}
\end{subequations}

In our theoretical approach we apply a Grad--Sonine (GS) methodology \cite{vNE98,G49}. First, the expansion \eqref{phi} is \emph{truncated} after $j+k+2\ell=2$ \cite{BB12}, so that the only retained coefficients are $a_{00}^{(0)}=1$ and those in Eqs.\ \eqref{ajkl}. Next, the collisional moments $\mu_{20}^{(0)}$, $\mu_{02}^{(0)}$, $\mu_{40}^{(0)}$, $\mu_{04}^{(0)}$, $\mu_{22}^{(0)}$, and $\mu_{00}^{(2)}$ are evaluated by inserting the truncated expansion into the collision operator $\mathcal{J}$, neglecting terms that are quadratic in the cumulants, and performing the velocity integrals. The resulting expressions (with coefficients being nonlinear functions of the temperature ratio $\theta$, the two coefficients of restitution $\alpha$ and $\beta$, and the dimensionless moment of inertia $\kappa\equiv {4I}/{m\sigma^2}$)  can be found in Ref.\ \cite{note_13_09_1}. Finally, the coupled set of five equations \eqref{partialtheta} and \eqref{partialpq} with $p+q+2r= 4$ are numerically solved to obtain the time evolution of $\theta$ and the cumulants \eqref{ajkl} \cite{note_13_09_1}. Setting $\partial_\tau\to 0$, the solution to the corresponding set of algebraic equations gives the stationary values of those quantities. To check the stability of the stationary values, we have analyzed the associated linearized problem and observed that all the eigenvalues have indeed a negative real part.
The characteristic relaxation period (in units of the accumulated number of collisions per particle) is $-1/\text{Re}(s)$, where $s$ is the eigenvalue with the real part closest to the origin. It {strongly increases} when the roughness parameter decreases from $\beta\approx 0$ to $\beta\gtrsim -1$ but is hardly dependent on the inelasticity parameter $\alpha$ \cite{note_13_09_1}.

\begin{figure}
\includegraphics[width=8cm]{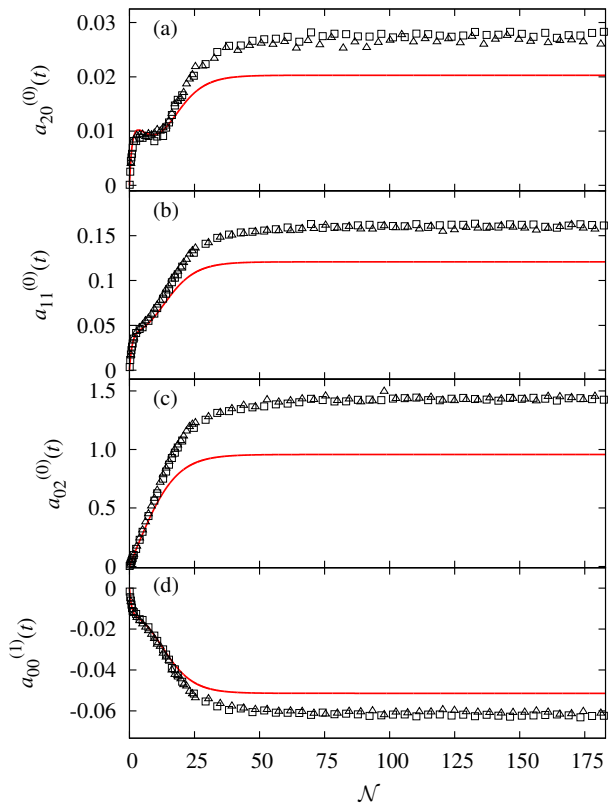}
\caption{(Color online) Temporal evolution  of the velocity {cumulants for} $\alpha=0.7$ and $\beta=-0.575$.  Henceforth, lines stand for theoretical results and symbols for {simulation data (DSMC: $\square$; MD: $\triangle$).}}
\label{fig1}
\end{figure}

In order to check the accuracy of our GS approximation, we have {performed DSMC and MD simulations} in the case of uniform spheres ($\kappa=\frac{2}{5}$), {starting from an initial equilibrium state, for a large number of $(\alpha,\beta)$ pairs}.  As a {particularly unfavorable case (see below)}, in Fig.\ \ref{fig1} we plot  the temporal evolution (as measured by the accumulated number of collisions per particle $\mathcal{N}$) of the cumulants for $\alpha=0.7$ and $\beta=-0.575$. As we see, {MD and DSMC results are hardly distinguishable, which reinforces the validity of the BE \eqref{BE} for dilute granular gases. Moreover,} both theory and simulation results agree very well in the first stages of development (up to $\mathcal{N}\approx 10$ collisions per particle). Beyond that stage, the angular velocity kurtosis  $a_{02}^{(0)}$ becomes larger than about ${0.3}$ and the GS theory (being based on truncation and linearization around the Maxwellian) underestimates the magnitude of the cumulants. However, the theory successfully captures the qualitative later evolution {and the duration of the total relaxation period}.
{The discrepancies between theory and simulations observed in Fig.\ \ref{fig1} for $\mathcal{N}\gtrsim 10$ are not due to an inherent limitation of the GS theory to the early stages of evolution but to the high values reached by the cumulant $a_{02}^{(0)}$ in this particularly stringent case. In fact, a good agreement is found at all times  for most combinations of $(\alpha,\beta)$  since in those cases the magnitudes of the cumulants are smaller than about ${0.3}$ \cite{note_13_09_1}}.

\begin{figure}
\includegraphics[width=8cm]{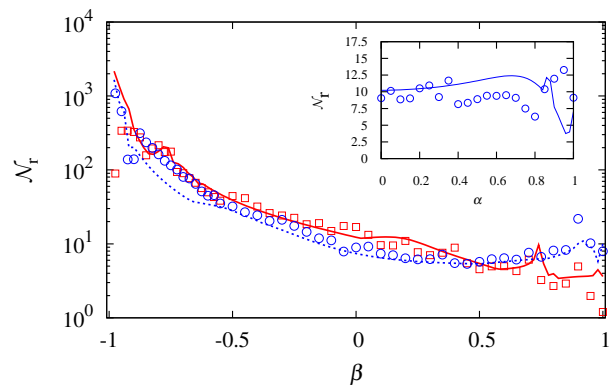}
\caption{(Color online) Relaxation time $\mathcal{N}_r$ (in units of collisions per particle)   as a function of $\beta$ for constant $\alpha=0.7$ (dotted line; {$\circ$}) or $\alpha=0.9$ (solid line; {$\square$}). The inset shows $\mathcal{N}_r$ as a function of $\alpha$ for constant $\beta=0$. {Symbols stand for DSMC data}.}
\label{fig2}
\end{figure}

To characterize the duration {($\mathcal{N}_r$)} of the relaxation period, we adopt the practical criterion
that the values of $\theta$ and $a_{jk}^{(\ell)}$ (with $j+k+2\ell =2$) must differ from their stationary values by less than $5\%$ if $\mathcal{N}>\mathcal{N}_r$. The theoretical and simulation results are presented in Fig.\ \ref{fig2}, where a good agreement is found. {Since each  quantity satisfies the $5\%$ criterion after a different relaxation period, what is plotted in Fig.\ \ref{fig2} is the maximum of the five particular relaxation times \cite{note_13_09_1}. This, together with the cases where  the relaxation is not monotonic, explains  the nonsmooth shape of $\mathcal{N}_r$ at some points.}
We may see that the $\alpha$ dependence is not as critical as the $\beta$ dependence, with $\mathcal{N}_r$ increasing dramatically as $\beta$ approaches the smooth limit $\beta\to -1$. This agrees with  the observed behavior of the theoretical quantity $-1/\text{Re}(s)$ \cite{note_13_09_1}. Interestingly, the inset of {Fig.\ \ref{fig2}} ({where the intermediate roughness $\beta=0$ is chosen as  a representative example}) {shows} that the relaxation time ($\mathcal{N}_r\approx 10$) to the hydrodynamic state  is not significantly different in the extreme limiting cases of complete inelasticity ($\alpha=0$) and complete elasticity ($\alpha=1$).
{We have checked by DSMC simulations that sixth- and eighth-degree moments relax over essentially the same time scale as the fourth-degree ones.}

\begin{figure}
\includegraphics[width=8cm]{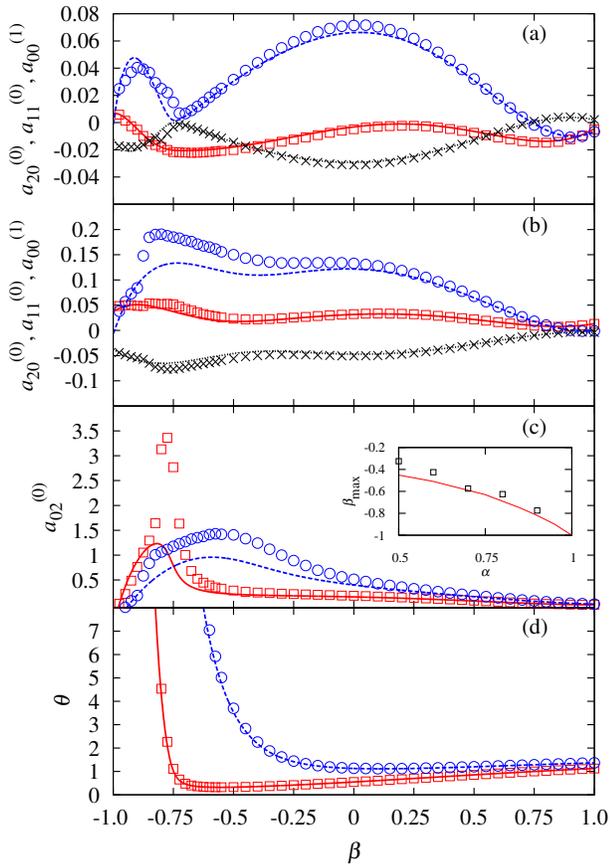}
\caption{(Color online) Stationary values of the cumulants and of the {temperature ratio} as functions of $\beta$. (a) ($\alpha=0.9$) and (b) ($\alpha=0.7$) show  $a_{20}^{(0)}$ (solid lines; {$\square$}),  $a_{11}^{(0)}$ (dashed lines; {$\circ$}), and $a_{00}^{(1)}$ (dotted lines, {$\times$}). (c) and (d) correspond to  $a_{02}^{(0)}$ and $\theta$, respectively,   for $\alpha=0.9$ (solid lines; {$\square$}) and $\alpha=0.7$ (dashed lines; {$\circ$}). The inset in (c) shows the value $\beta_{\text{max}}(\alpha)$ at which $a_{02}^{(0)}$ reaches its maximum value. {Symbols stand for DSMC data}.}
\label{fig3}
\end{figure}

Let us now focus on the hydrodynamic (``steady'') states.  In Fig.\ \ref{fig3} we plot the cumulants and the temperature ratio as functions of $\beta$ for $\alpha=0.7$ and $0.9$. The agreement between theory and simulation for the translational velocity kurtosis $a_{20}^{(0)}$, the (orientational) translational-rotational  correlation parameter $a_{00}^{(1)}$, and the temperature ratio $\theta$ is excellent for all values of $\beta$, especially at $\alpha=0.9$. The agreement for $a_{11}^{(0)}$ is still very good, except in the ranges where $a_{02}^{(0)}$ reaches  high values ($a_{02}^{(0)}\gtrsim {0.3}$). Note that, according to our simulations,  the angular velocity kurtosis  at $\alpha=0.9$ reaches values as high as $a_{02}^{(0)}\approx 3$ if the roughness is tuned to $\beta= \beta_{\text{max}}\simeq -0.78$. As the inelasticity increases to $\alpha=0.7$, the maximum of $a_{02}^{(0)}$ decreases to $a_{02}^{(0)}\approx 1.5$ and occurs at a larger roughness ($\beta_{\text{max}}\simeq -0.58$). {The latter} is precisely the case analyzed in Fig.\ \ref{fig1}. As seen from the inset in Fig.\ \ref{fig3}(c), the GS approximation captures qualitatively well the $\alpha$-dependence of $\beta_{\text{max}}$. {The existence of very large values of $a_{02}^{(0)}$ at $(\alpha,\beta_{\max})$ does not  have a straightforward intuitive explanation but shows a subtle interplay between the translational and rotational degrees of freedom in the granular gas.}

\begin{figure}
\includegraphics[width=8cm]{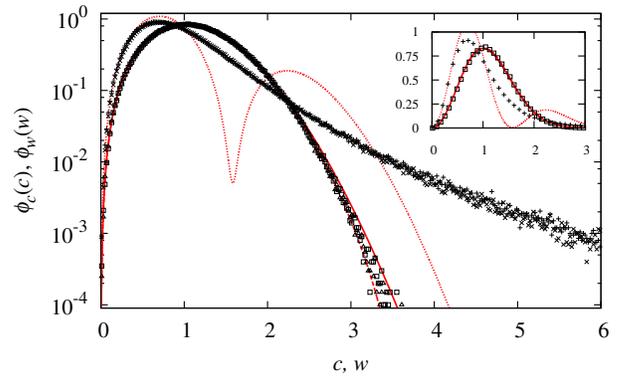}
\caption{(Color online) Marginal distribution functions $\phi_c(c)$ {(dashed line; DSMC: $\square$;  MD: $\triangle$)} and $\phi_w(w)$ {(dotted line; DSMC: $+$;  MD: $\times$)} at $\alpha=0.9$, {$\beta=-0.75$}. The solid line stands for the (common) Maxwellian distribution function.}
\label{fig4}
\end{figure}

The results of Fig.\ \ref{fig3} show that the hydrodynamic HCS distribution function can be highly non-Maxwellian, especially with respect to the angular velocities, if at a given $\alpha$ the roughness parameter $\beta$ is close to $\beta_{\text{max}}(\alpha)$. In those cases, {truncation of the perturbative expansion \eqref{phi} is not the adequate tool to accurately describe the distribution function, regardless of the truncation order, and an alternative approach is needed.}
This is illustrated by Fig.\ \ref{fig4}, where  the marginal distribution functions $\phi_c(c)=4\pi c^2\int d\mathbf{w}\phi(\mathbf{c},\mathbf{w})$ and $\phi_w(w)=4\pi w^2\int d\mathbf{c}\phi(\mathbf{c},\mathbf{w})$ are plotted for the extreme case (see Fig.\ \ref{fig3}) $\alpha=0.9$, {$\beta\gtrsim\beta_{\text{max}}=-0.75$}. While $\phi_c(c)$ is close to the Maxwellian and is well represented by the GS truncated expansion, a large discrepancy is observed between the actual distribution $\phi_w(w)$ and the corresponding GS distribution, even if the latter is parametrized with the empirical kurtosis $a_{02}^{(0)}$. The GS distribution $\phi_w(w)$ is bimodal (with an almost zero local minimum) whereas simulation data do not {exhibit} this feature. More interestingly, {both DSMC and MD} simulation data for $\phi_w(w)$ show extremely large high-energy tails (perhaps the largest ones reported for granular gases of hard spheres {so far \cite{GA08}}) consistent with  $\phi_w(w)\sim \exp(-A w)$.

In summary, we have studied the temporal evolution of the HCS for a granular gas of rough hard spheres by a GS truncated expansion and by DSMC {and MD} simulations. {The three} methods confirm that, after a kinetic stage, a hydrodynamic regime is reached where the whole time dependence of the velocity distribution function is enslaved by the temperature. The GS theory provides an excellent description of the evolution of the temperature ratio and the four velocity cumulants, except when the angular velocity kurtosis   becomes {so large} ($a_{02}^{(0)}\gtrsim {0.3}$) {that it} compromises the assumptions behind the truncation and linearization scheme. Even in those cases, the GS theory predicts well the relaxation time (see Fig.\ \ref{fig2}) and  describes qualitatively  the roughness dependence of the stationary cumulants (see Fig.\ \ref{fig3}). Quite {surprisingly, and in contrast to what was generally believed}, the relaxation to the hydrodynamic state is {practically independent of} the inelasticity coefficient $\alpha$ and is rather fast (no more than about ten collisions per particle, even in the most inelastic case, $\alpha=0$) if the spheres are sufficiently rough {($\beta\gtrsim 0$)}. Therefore, we may conclude that high inelasticity does not preclude by itself the applicability of hydrodynamics. On the other hand, {paradoxically}, if the spheres are weakly rough ($\beta\gtrsim -1$), the relaxation time  increases dramatically to values on the order of at least $10^3$ collisions per particle. This is because, as roughness decreases, more and more collisions are needed to activate the rotational degrees of freedom, which are absolutely {quenched} in the smooth-sphere model.
{Interestingly}, the duration of the relaxation stage and the departure from the (two-temperature) Maxwellian are not fully correlated, as comparison between Figs.\ \ref{fig2} and \ref{fig3} shows. In particular, at a given $\alpha$, the maximum distortion from the Maxwellian (as monitored by the kurtosis $a_{02}^{(0)}$) does not take place in the limit $\beta\to -1$ but at a certain value $\beta_{\text{max}}(\alpha)<0$ [see the inset in Fig.\ \ref{fig3}(c)].

Given that the HCS is the base state for a granular gas and for the application of the Chapman--Enskog method \cite{D01}, we expect these results to be of help in further developments of hydrodynamic transport theories of inhomogeneous granular gases.
In this respect, it is interesting to note that most of the materials are characterized by positive values of the roughness parameter (typically, $\beta\sim 0.5$) \cite{L99},
where the GS theory developed here is highly accurate.

\acknowledgments
F.V.R. and A.S. acknowledge support of the Spanish Government through Grant No.\ FIS2010-16587 and
the Junta de Extremadura (Spain) through Grant No.\ GR10158, partially
financed by Fondo Europeo de Desarrollo Regional (FEDER) funds. The work by of G.M.K. has been supported by the Conselho Nacional de Desenvolvimento Cient\'ifico e Tecnol\'ogico (Brazil).

\bibliographystyle{apsrev}

\bibliography{D:/Dropbox/Public/bib_files/Granular}

\begin{thebibliography}{33}
\bibitem[{not({\natexlab{a}})}]{note_13_11_1}
\bibinfo{note}{We used $N=7700$ particles in a cubic box (with periodic
  boundary conditions) of side $L=400\sigma$, which corresponds to a volume
  fraction $\varphi\simeq 6.3\times 10^{-5}$. No clustering instability was
  observed under these conditions. Since granular temperatures decrease
  monotonically in time, particle velocities were rescaled periodically in
  order to avoid numerical errors due to exceedingly small magnitudes.}
\end{thebibliography}

\appendix

\section{Supplemental material}

{In this supplemental material we give  the reduced collision operator $\mathcal{J}$,   the orthogonality relation for the polynomials $\Psi_{jk}^{(\ell)}$,  the explicit expressions of the reduced collisional moments $\mu_{jk}^{(\ell)}$ (with $j+k+2\ell\leq 4$) in the Grad--Sonine approximation, and additional curves comparing theoretical and simulation results.}

\subsection{Collision operator}
{The collision operator $K[\mathbf{v}, \bm{\omega}|f]$ introduced in Eq.\ (1) of the main paper is}
\begin{eqnarray}
  \label{CopA}
K[\mathbf{v}_1, \bm{\omega}_1|f]&  =& \int d\mathbf{v}_2\int d{\bm\omega}_2\int d\widehat{\bm{\sigma}}\,\Theta(\mathbf{v}_{12}\cdot\widehat{\bm{\sigma}})(\mathbf{v}_{12}\cdot\widehat{\bm{\sigma}}) \nn
& &\times\left[(\alpha\beta)^{-2}f_1''f_2''-f_1f_2\right].
\end{eqnarray}
Here,  $\Theta$ is the Heaviside step function, $\mathbf{v}_{12}=\mathbf{v}_1-\mathbf{v}_2$ is the relative velocity, and $\widehat{\bm{\sigma}}$ is the unit vector pointing from the center of sphere 1 to the center of sphere 2. Also,  $f_i\equiv f(\mathbf{v}_i,\bm{\omega}_i,t)$, $f_i''\equiv f(\mathbf{v}_i'',\bm{\omega}_i'',t)$, where the double primes indicate pre-collisional velocities.

In terms of the reduced quantities
\beq
\mathbf{c}\equiv\frac{\mathbf{v}-\mathbf{u}_f}{\sqrt{2T_t/m}}, \quad
\mathbf{w}\equiv\frac{\bm{\omega}}{\sqrt{2T_r/I}},
\eeq
\beq
\phi(\mathbf{c},\mathbf{w})\equiv \frac{1}{n}\left(\frac{4T_tT_r}{m I}\right)^{3/2}f(\mathbf{v},\bm{\omega}),
\eeq
the reduced collision operator $\mathcal{J}$    reads
\begin{eqnarray}
  \label{Cop}
\mathcal{J}[\mathbf{c}_1, \mathbf{w}_1|\phi]&  =&\frac{1}{\sqrt{2\pi}} \int d\mathbf{c}_2\int d\mathbf{w}_2\int d\widehat{\bm{\sigma}}\,\Theta(\mathbf{c}_{12}\cdot\widehat{\bm{\sigma}})(\mathbf{c}_{12}\cdot\widehat{\bm{\sigma}}) \nn
& &\times\left[(\alpha\beta)^{-2}\phi_1''\phi_2''-\phi_1\phi_2\right].
\end{eqnarray}
The  pre-collisional velocities are given by the restituting collision rules \cite{SKS11}
\beq
\mathbf{c}_{1}''=\mathbf{c}_{1}-\JJb, \quad \mathbf{c}_{2}''=\mathbf{c}_{2}+\JJb,
\label{25}
\eeq
\beq
\mathbf{w}_1''=\mathbf{w}_1-\frac{1}{\sqrt{\kappa\theta}}\widehat{\bm{\sigma}}\times\JJb,
\quad \mathbf{w}_2''=\mathbf{w}_2-\frac{1}{\sqrt{\kappa\theta}}\widehat{\bm{\sigma}}\times\JJb,
\label{26}
\eeq
where
\beqa
\JJb&=&\frac{\at}{\al}\left(\mathbf{c}_{12}\cdot\widehat{\bm{\sigma}}\right)\widehat{\bm{\sigma}}+
\frac{\bt}{\beta}\Bigg[\mathbf{c}_{12}-\left(\mathbf{c}_{12}\cdot\widehat{\bm{\sigma}}\right)\widehat{\bm{\sigma}}\nn
&&
-\sqrt{\frac{\theta}{\kappa}}\widehat{\bm{\sigma}}\times\left(\mathbf{w}_1+\mathbf{w}_2\right)\Bigg].
\label{27}
\eeqa
Here,
\beq
\at\equiv\frac{1+\al}{2},\quad\bt\equiv\frac{\q}{1+\q}\frac{1+\be}{2}.
\label{20a}
\eeq
For the sake of completeness, we also give the post-collisional velocities (direct collision rules):
\beq
\mathbf{c}_{1}'=\mathbf{c}_{1}-\JJ, \quad \mathbf{c}_{2}'=\mathbf{c}_{2}+\JJ,
\label{25b}
\eeq
\beq
\mathbf{w}_1'=\mathbf{w}_1-\frac{1}{\sqrt{\kappa\theta}}\widehat{\bm{\sigma}}\times\JJ,
\quad \mathbf{w}_2'=\mathbf{w}_2-\frac{1}{\sqrt{\kappa\theta}}\widehat{\bm{\sigma}}\times\JJ,
\label{26b}
\eeq
\beqa
\JJ&=&\at\left(\mathbf{c}_{12}\cdot\widehat{\bm{\sigma}}\right)\widehat{\bm{\sigma}}+
\bt\Bigg[\mathbf{c}_{12}-\left(\mathbf{c}_{12}\cdot\widehat{\bm{\sigma}}\right)\widehat{\bm{\sigma}}\nn
&&
-\sqrt{\frac{\theta}{\kappa}}\widehat{\bm{\sigma}}\times\left(\mathbf{w}_1+\mathbf{w}_2\right)\Bigg].
\label{27b}
\eeqa

\subsection{Orthogonality relation for the polynomials $\Psi_{jk}^{(\ell)}$}
The complete set of polynomials for the expansion of $\phi(\mathbf{c},\mathbf{w})/\phi_M(\mathbf{c},\mathbf{w})$ is defined by
\beq
\Psi_{jk}^{(\ell)}(\mathbf{c},\mathbf{w})=L_j^{(2\ell+\frac{1}{2})}(c^2)L_k^{(2\ell+\frac{1}{2})}(w^2)\left(c^2w^2\right)^\ell P_{2\ell}(u).
\label{Psi}
\eeq
The orthogonality relations of the Laguerre and Legendre polynomials are
\beq
\int_0^\infty dx\, e^{-x} x^{\alpha}L_j^{(\alpha)}(x)L_{j'}^{(\alpha)}(x)=\frac{\Gamma(j+\alpha+1)}{nj}\delta_{j,j'},
\label{ortho1}
\eeq
\beq
\int_0^1 dx\, P_{2\ell}(x)P_{2\ell'}(x)=\frac{1}{4\ell+1}\delta_{\ell,\ell'}.
\label{ortho2}
\eeq
We will also need the property
\beq
\int_0^\infty dx\, e^{-x} x^{\alpha-1}L_j^{(\alpha)}(x)={\Gamma(\alpha)}.
\label{ortho3}
\eeq

Let us define the scalar product of two arbitrary isotropic (real) functions $\Phi(\mathbf{c},\mathbf{w})$ and $\Phi'(\mathbf{c},\mathbf{w})$ as
\beqa
\langle \Phi|\Phi'\rangle&\equiv& \int d\mathbf{c}\int d\mathbf{w} \,\phi_M(\mathbf{c},\mathbf{w}) \Phi(\mathbf{c},\mathbf{w})\Phi'(\mathbf{c},\mathbf{w})\nn
&=&
\frac{4}{\pi}\int_0^\infty dc^2\, e^{-c^2}c\int_0^\infty dw^2\, e^{-w^2}w\int_0^1 du\nn
&&\times\Phi(\mathbf{c},\mathbf{w})\Phi'(\mathbf{c},\mathbf{w}),
\eeqa
where in the last step we have taken into account that $\int d\mathbf{c}\int d\mathbf{w}\to 4\pi^2\int_0^\infty dc^2\, c\int_0^\infty dw^2\, w\int_0^1 du$.
Then,
\beqa
\langle \Phi\rangle&=&\int d\mathbf{c}\int d\mathbf{w}\, \Phi(\mathbf{c},\mathbf{w})\phi(\mathbf{c},\mathbf{w})\nn
&=&\left\langle \Phi|{\phi}/{\phi_M}\right\rangle.
\eeqa

The orthogonality properties \eqref{ortho1} and \eqref{ortho2} imply
\beq
\langle \Psi_{jk}^{(\ell)}|\Psi_{j'k'}^{(\ell')}\rangle=N_{jk}^{(\ell)}\delta_{j,j'}\delta_{k,k'}\delta_{\ell,\ell'},
\label{avPsi}
\eeq
where
\beq
N_{jk}^{(\ell)}\equiv \frac{\Gamma(j+2\ell+\frac{3}{2})\Gamma(k+2\ell+\frac{3}{2})}{[\Gamma(\frac{3}{2})]^2(4\ell+1)j!k!}.
\eeq
As a consequence, the coefficients in the expansion (6) of the main paper are
\beq
a_{jk}^{(\ell)}=\frac{\langle\Psi_{jk}^{(\ell)}\rangle}{N_{jk}^{(\ell)}} .
\eeq

\subsection{Collisional moments\label{sec3b}}
The truncation of the expansion (6) of the main paper after $j+k+2\ell\geq 3$ gives
\begin{widetext}
\beq
\frac{\phi(\mathbf{c},\mathbf{w})}{\phi_M(\mathbf{c},\mathbf{w})}\approx 1+{a_{20}^{(0)}}\frac{{15}-20c^2+4c^4}{8}+{a_{02}^{(0)}}\frac{{15}-20w^2+4w^4}{8}+{a_{11}^{(0)}} \frac{\left(3-2c^2\right)\left(3-2w^2\right)}{4}
+{a_{00}^{(1)}}\frac{3(\mathbf{c}\cdot\mathbf{w})^2-c^2w^2}{2}.
\label{20}
\eeq

By inserting the approximation \eqref{20} into Eq.\ \eqref{Cop} and neglecting terms nonlinear in $a_{20}^{(0)}$, $a_{11}^{(0)}$, $a_{02}^{(0)}$, and  $a_{00}^{(1)}$, we have  obtained the following expressions for the second- and fourth-degree  collisional moments:
\beq
{\mu_{20}^{(0)}}={4\left[\at\left(1-\at\right)+\bt\left(1-\bt\right)\right]\left(1+\frac{3a_{20}^{(0)}}{16}\right)}{-\frac{4\bt^2\theta}{\kappa}\left(1-\frac{a_{20}^{(0)}}{16}+\frac{2a_{11}^{(0)}{-a_{00}^{(1)}}}{8}\right)},
\label{22}
\eeq
\beq
\mu_{02}^{(0)}= \frac{4\bt}{\kappa}\left[\left(1-\frac{\bt}{\kappa}\right)\left(1-\frac{a_{20}^{(0)}}{16}+
\frac{2a_{11}^{(0)}{-a_{00}^{(1)}}}{8}\right)-\frac{\bt}{\theta}\left(1+\frac{3a_{20}^{(0)}}{16}\right)\right].
\label{23}
\eeq
\beqa
\mu_{40}^{(0)}&=&16\left[\at^3\left(2-\at\right)+\bt^3\left(2-\bt\right)-\at\bt \left(1-\at-\bt+\at\bt\right)\right]+22\left(\at+\bt\right)-38\left(\at^2+\bt^2\right)\nn
&&-15\left[\at\bt \left(\frac{23}{15}-\at-\bt+\at\bt\right)-\frac{269}{120}\left(\at+\bt\right)+\frac{357}{120}\left(\at^2+\bt^2\right)-\at^3\left(2-\at\right)-\bt^3\left(2-\bt\right)\right]
a_{20}^{(0)}\nn
&&-\frac{22 \bt^2\theta}{\kappa}\left(1+\frac{41a_{20}^{(0)}}{176 }+3\frac{2a_{11}^{(0)}{-a_{00}^{(1)}}}{8}\right)+\frac{16\bt^2\theta}{\kappa}\left[\at\left(1-\at\right)+2\bt\left(1-\bt\right)\right]
\left(1+\frac{3a_{20}^{(0)}}{16}+3\frac{2a_{11}^{(0)}{-a_{00}^{(1)}}}{8}\right)\nn
&&-\frac{16\bt^4 \theta^2}{\kappa^2}\left(1-\frac{a_{20}^{(0)}}{16}+\frac{a_{02}^{(0)}}{2}+\frac{2a_{11}^{(0)}{-a_{00}^{(1)}}}{4}\right),
\label{28}
\eeqa
\beqa
\mu_{22}^{(0)}&=&
6\left[\at\left(1-\at\right)+\bt\left(1-\bt\right)-\frac{4\at\bt}{3\kappa}\left(1-\at\right)\left(1-\frac{\bt}{\kappa}\right)-\frac{8\bt^2}{3\kappa}
\left(\frac{3}{4}-\bt-\frac{\bt}{\kappa}+2\frac{\bt^2}{\kappa}\right)\right]\left(1+\frac{3a_{20}^{(0)}}{16}+3\frac{2a_{11}^{(0)}{-a_{00}^{(1)}}}{8}\right)
\nn
&&+\frac{7\bt}{\kappa}
\left(1-\frac{\bt}{\kappa}\right)\left(1+\frac{29a_{20}^{(0)}}{112}\right)-\frac{3\bt^2}{2\kappa\theta}a_{20}^{(0)}
-\frac{8\bt^2}{\kappa\theta}\Bigg[\frac{9}{8}-\at\left(1-\at\right)-2\bt\left(1-\bt\right)
\Bigg]\left(1+\frac{15a_{20}^{(0)}}{16}\right)\nn
&&-\frac{\bt^2\theta}{\kappa}\left[5-8 \frac{\bt}{\kappa}\left(1-\frac{\bt}{\kappa}\right)\right]a_{02}^{(0)}-\frac{8\bt^2\theta}{\kappa}\left[1-2 \frac{\bt}{\kappa}\left(1-\frac{\bt}{\kappa}\right)\right] \left(1-\frac{a_{20}^{(0)}}{16}+\frac{2a_{11}^{(0)}{-a_{00}^{(1)}}}{4}\right)\nn
&&+3\left[\frac{\bt}{\kappa}
\left(\frac{37}{12}-2\bt-\frac{7\bt}{4\kappa}\right)+\at+\bt-\frac{4 \at\bt}{3\kappa}\right]\frac{2a_{11}^{(0)}{-a_{00}^{(1)}}}{2}{+\Bigg[5\left(\at+\bt\right)-3\left(\at^2+\bt^2\right)}
{+\frac{4\bt}{\q}\left(1-\bt\right)}\nn
&&{-\frac{\bt^2}{\q^2}\left(2+\q\theta\right)\Bigg] \frac{3a_{00}^{(1)}}{4}},
\label{29}
\eeqa
\beqa
\mu_{04}^{(0)}&=&\frac{\bt}{\kappa}\left\{4\left(1-\frac{\bt}{\kappa}\right)\left[5-4\frac{\bt}{\kappa}
\left(1-\frac{\bt}{\kappa}\right)\right]\left(1-\frac{a_{20}^{(0)}}{16}\right)- \frac{4\bt}{\theta}\left[5-8\frac{\bt}{\kappa}\left(1-\frac{\bt}{\kappa}\right)\right]
\left(1+\frac{3a_{20}^{(0)}}{16}+3\frac{2a_{11}^{(0)}{-a_{00}^{(1)}}}{8}\right)
\right.\nn
&&-{5}\left(1-\frac{4 \bt}{5\kappa}\right)\left({2a_{11}^{(0)}{-a_{00}^{(1)}}}\right)-\frac{16\bt^3}{\kappa\theta^2}\left(1+\frac{15a_{20}^{(0)}}{16}\right)
+4
\left(5-\frac{13}{2}\frac{\bt}{\kappa}+4\frac{\bt^2}{\kappa^2} - 2\frac{\bt^3}{\kappa^3}\right)\left(a_{02}^{(0)}+\frac{2a_{11}^{(0)}{-a_{00}^{(1)}}}{2}\right)\nn
&&\left.
{+\left(1-\frac{\bt}{\kappa}-\frac{\bt}{\theta}\right)\frac{3a_{00}^{(1)}}{2}}\right\},
\label{30}
\eeqa
\beqa
{\mu_{00}^{(2)}}&=&{2\left[\at\left(1-\at\right)-\bt^2\left(1+\frac{1}{\q^2}\right)\right]
\left(1+\frac{3a_{20}^{(0)}}{16}+\frac{3a_{11}^{(0)}}{4}+\frac{3a_{00}^{(1)}}{4}\right)
+{\at}\left(a_{11}^{(0)}+4a_{00}^{(1)}\right)
+2\bt\left(1+\frac{1-\bt}{\q}\right)}\nn
&&\times\left(1+\frac{3a_{20}^{(0)}}{16}+\frac{5a_{11}^{(0)}}{4}+\frac{13a_{00}^{(1)}}{8}\right)
+3\bt\left(\frac{3}{4}-\at\right)\left(1+\frac{1}{\q}\right){a_{00}^{(1)}}-\frac{\bt^2}{\q\theta}\left(1+\frac{7a_{20}^{(0)}}{16}\right)
\nn
&&-\frac{\bt^2\theta}{\q}\left(1-\frac{a_{20}^{(0)}}{16}+\frac{2a_{11}^{(0)}{-a_{00}^{(1)}}}{4}\right).
\label{30n}
\eeqa

\end{widetext}

{Equations \eqref{22}--\eqref{30} generalize the results derived in Ref.\ \cite{SKS11} by including the terms associated with  $a_{00}^{(1)}\neq 0$. Furthermore, Eq.\ \eqref{30n} generalizes the results of Refs.\ \cite{BPKZ07,KBPZ09} by considering $a_{20}^{(0)}\neq 0$, $a_{11}^{(0)}\neq 0$, and $a_{02}^{(0)}\neq 0$. As an additional simple consistency test, we get $\mu_{22}^{(0)}=3\mu_{00}^{(2)}=\frac{3}{2}\mu_{20}^{(0)}=
\frac{3}{2}(1-\alpha^2)\left(1+\frac{3}{16}a_{20}^{(0)}\right)$ in the special case of smooth spheres ($\be=-1$) with $a_{11}^{(0)}=a_{00}^{(1)}=0$.
This condition means that, if the inelastic spheres are perfectly smooth (so the translational and rotational degrees of freedom are totally uncoupled), the collisional rates of change of $c^2w^2$ and
$(\mathbf{c}\cdot\mathbf{w})^2$ are $\langle w^2\rangle=\frac{3}{2}$ and $\langle w_z^2\rangle=\frac{1}{2}$ times, respectively, the collisional rate of change of $c^2$.}

\begin{figure}[tbp]
\includegraphics[width=7cm]{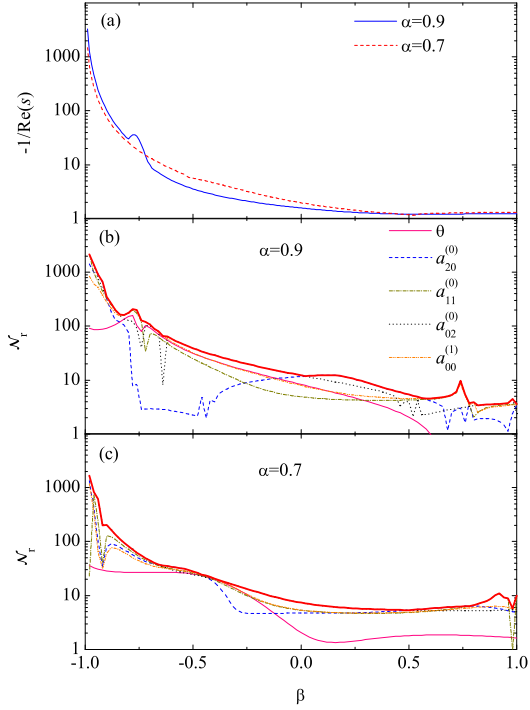}
\caption{(a) Plot of $-1/\text{Re}(s)$, where $s$ is the eigenvalue with the real part closest to the origin, as a function of $\beta$ for $\alpha=0.9$ and $0.7$. (b) Relaxation periods $\mathcal{N}_r$ corresponding to $\theta$ and to the fourth-degree cumulants as functions of $\beta$ for $\alpha=0.9$. The tick solid line is the upper envelope of the individual curves. (c) Same as in (b) but for $\alpha=0.7$.}
\label{fig0A}
\end{figure}

\begin{figure}[tbp]
\includegraphics[width=7.cm]{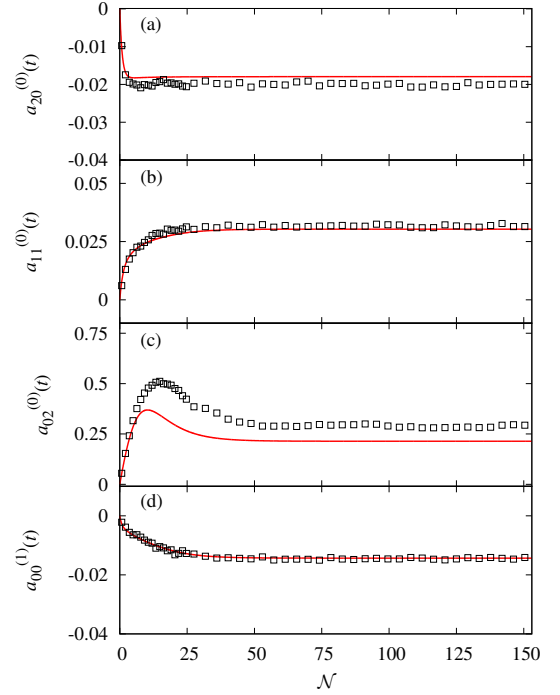}
\caption{Temporal evolution  of the velocity cumulants (a) $a_{20}^{(0)}$, (b) $a_{11}^{(0)}$, (c) $a_{02}^{(0)}$, and (d) $a_{00}^{(1)}$, for $\alpha=0.9$ and $\beta=-0.5$.  Lines stand for theoretical results and symbols for DSMC simulations.}
\label{fig1A}
\end{figure}
\begin{figure}[tbp]
\includegraphics[width=7.cm]{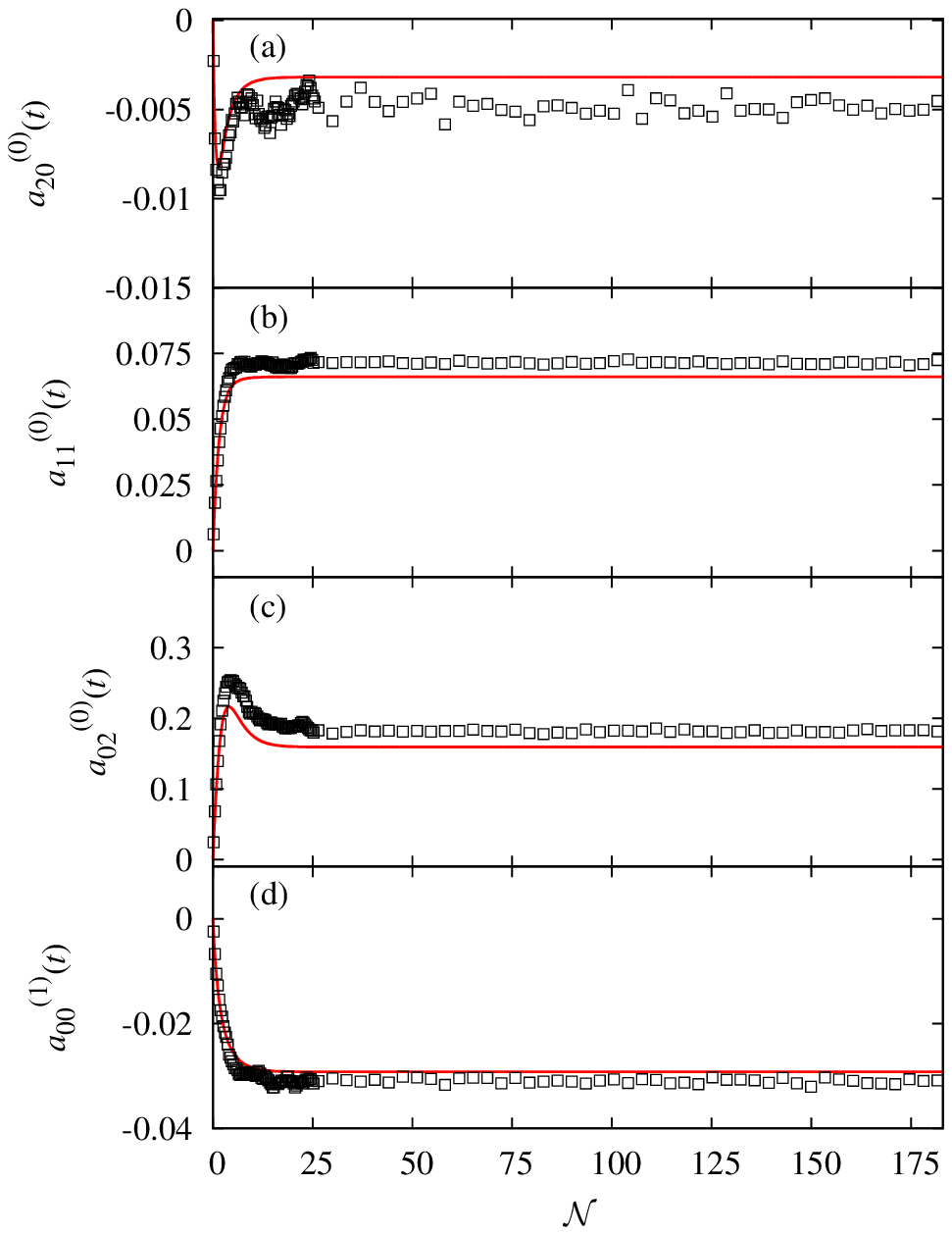}
\caption{Same as in Fig.\ \protect\ref{fig1A} but for $\alpha=0.9$ and $\beta=0$.}
\label{fig2A}
\end{figure}
\begin{figure}[tbp]
\includegraphics[width=7.cm]{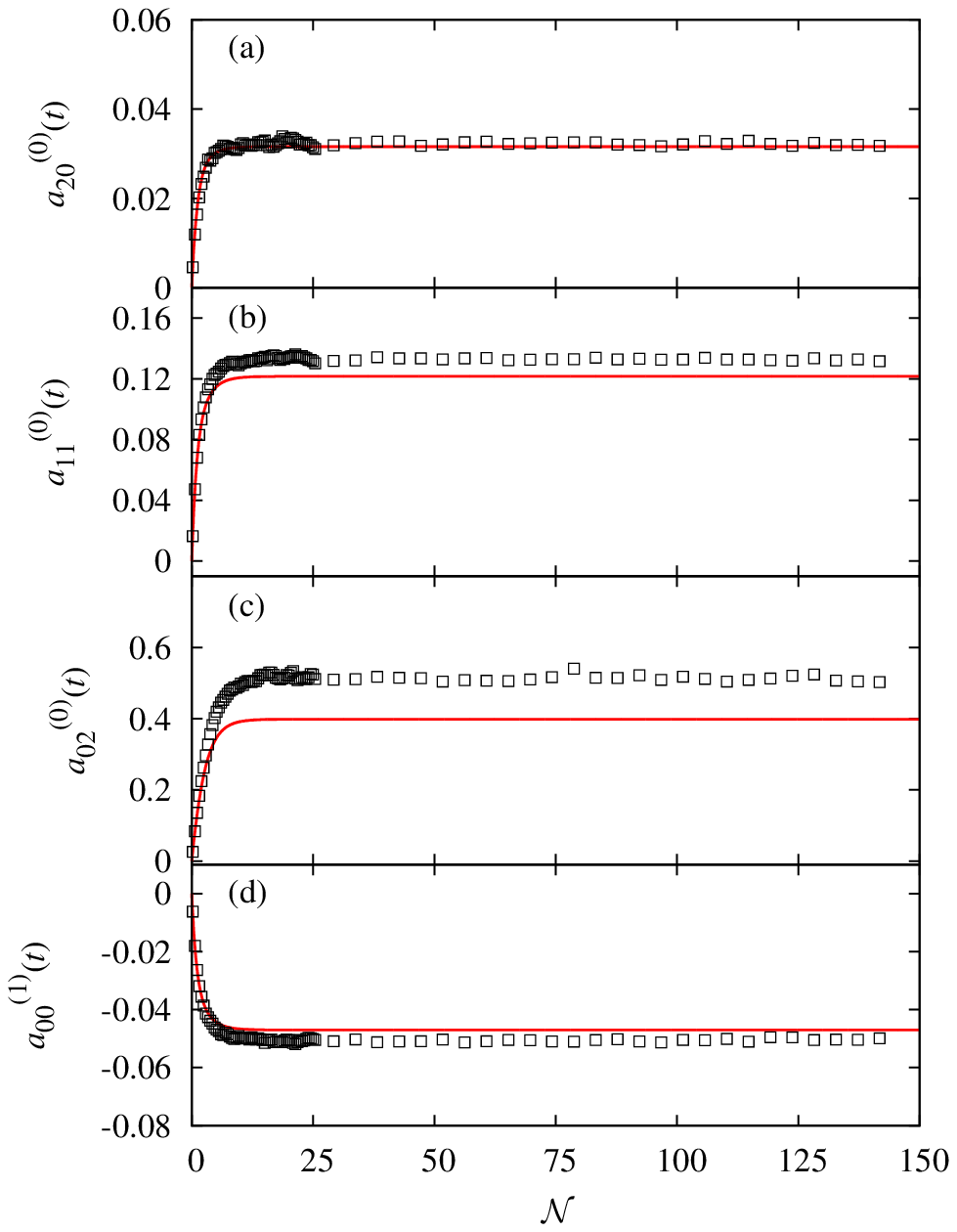}
\caption{Same as in Fig.\ \protect\ref{fig1A} but for $\alpha=0.7$ and $\beta=0$.}
\label{fig3A}
\end{figure}

\subsection{Evolution equations}
{}From Eqs.\ (2) and (5) of the main paper one finds
\beq
\partial_\tau \ln\theta=\frac{2}{3}\mathcal{L}\left(\mu_{20}^{(0)}-\mu_{02}^{(0)}\right),
\label{evotheta}
\eeq
\beq
\partial_\tau \ln\left(1+a_{20}^{(0)}\right) =\frac{4}{15}\mathcal{L}\left(5\mu_{20}^{(0)}-\frac{\mu_{40}^{(0)}}{1+a_{20}^{(0)}}\right),
\label{evoa20}
\eeq
\beq
\partial_\tau \ln\left(1+a_{02}^{(0)}\right) =\frac{4}{15}\mathcal{L}\left(5\mu_{02}^{(0)}-\frac{\mu_{04}^{(0)}}{1+a_{02}^{(0)}}\right),
\label{evoa02}
\eeq
\beq
\partial_\tau \ln\left(1+a_{11}^{(0)}\right) =\frac{4}{9}\mathcal{L}\left(\frac{3}{2}\mu_{20}^{(0)}+\frac{3}{2}\mu_{02}^{(0)}-\frac{\mu_{22}^{(0)}}{1+a_{11}^{(0)}}\right),
\label{evoa11}\eeq
\beqa
\partial_\tau \ln\left(1+a_{11}^{(0)}+\frac{5}{2}a_{00}^{(1)}\right) &=&\frac{4}{3}\mathcal{L}\Bigg(\frac{1}{2}\mu_{20}^{(0)}+\frac{1}{2}\mu_{02}^{(0)}\nn
&&-\frac{\mu_{00}^{(2)}}{1+a_{11}^{(0)}+\frac{5}{2}a_{00}^{(1)}}\Bigg).
\label{evob}
\eeqa
Here, $\mathcal{L}(X)$ means that the quantity $X$ is linearized with respect to the cumulants (but not with respect to the temperature ratio $\theta$), so that Eqs.\ \eqref{evotheta}--\eqref{evob} constitute a closed set of equations.


{The stationary solutions are found by equating to zero the right-hand sides of Eqs.\ \eqref{evotheta}--\eqref{evob}. Since they are linear in the cumulants, Eqs.\ \eqref{evotheta}, \eqref{evoa20}, \eqref{evoa11}, and \eqref{evob} are first used to express the four cumulants as functions of $\theta$. Inserting those expressions into Eq.\ \eqref{evoa02} allows one to obtain a closed tenth-degree equation for $\theta$. The physical root is determined as the one close to the temperature ratio in the Maxwellian approximation, namely \cite{SKS11,KBPZ09}}
\beq
\theta\approx\sqrt{1+C^2}+C
\label{1.8}
\eeq
with
\beq
C\equiv \frac{1+\kappa}{2\kappa(1+\beta)}\left[(1+\kappa)\frac{1-\alpha^2}{1+\beta}-(1-\kappa)(1-\beta)\right].
\label{1.9}
\eeq

{The linear stability of the stationary values can be studied from the eigenvalues of the linearized set of evolution equations. We have observed that all the eigenvalues have a negative real part.
Figure \ref{fig0A}(a) shows the $\beta$-dependence of $-1/\text{Re}(s)$, where $s$ is the eigenvalue with the real part closest to the origin, for $\alpha=0.9$ and $0.7$.
The relaxation period {($\mathcal{N}_r$)} for a quantity $X$ is defined by the condition that $|(X-X_s)/X_s|<0.05$  during the subsequent evolution, where $X_s$ is the stationary value of $X$. The five individual relaxation times, along with their upper envelope, are plotted in Figs.\ \ref{fig0A}(b) and  \ref{fig0A}(c) for $\alpha=0.9$ and $0.7$, respectively. Depending on the values of $(\alpha,\beta)$ the longest relaxation time  corresponds to $a_{20}^{(0)}$, $a_{02}^{(0)}$, or $a_{11}^{(0)}$. The  quantities $\theta$ and $a_{00}^{(1)}$ typically relax earlier than at least one of the other three quantities.
The non-smooth shape at some points of the curve $\mathcal{N}_r$ associated with a certain quantity $X$  is a consequence of a non-monotonic relaxation. In that case, $X$ crosses  $X_s$  at a given time $\mathcal{N}=\mathcal{N}_0$, then  $|(X-X_s)/X_s|$ reaches a local maximum at $\mathcal{N}=\mathcal{N}_{\max}$, and finally $|X-X_s|\to 0$. If the value of $|(X-X_s)/X_s|$ at $\mathcal{N}=\mathcal{N}_{\max}$ is larger than $0.05$, then $\mathcal{N}_r>\mathcal{N}_{\max}$. Otherwise, $\mathcal{N}_r<\mathcal{N}_0$. Thus, a small change of $\alpha$ or $\beta$ can produce a much larger change in $\mathcal{N}_r$.}

\subsection{Additional simulation curves}
{Figures \ref{fig1A}--\ref{fig3A} compare the temporal evolution predicted by the GS approximation with that obtained from DSMC simulations for $(\alpha,\beta)=(0.9,-0.5)$, $(0.9,0)$, and $(0.7,0)$, respectively. As we see, the general agreement is much better than in the case displayed in Fig.\ 1 of the main paper.}

\end{document}